\documentclass[proceedings]{JHEP} 
\usepackage{epsfig}

\title{Review of Charm Lifetimes}
\author{Harry W.~K.~Cheung\thanks{Talk presented at the
8th International Symposium on Heavy Flavour Physics,
Southampton, U.K., 25--29 July 1999. This is FERMILAB-Conf-99/344.
This work was supported
by the Fermi National Accelerator Laboratory, which is
operated by the Universities Research Association, Inc.,
under contract DE-AC02-76CHO3000 with the U.S. Department
of Energy}\\
Fermi National Accelerator Laboratory, 
P.~O.~Box~500, Batavia, IL 60510-0500, U.S.A.\\
E-mail: \email{cheung@fnal.gov}}

\conference{Heavy Flavours 8, Southampton, UK, 1999}

\abstract{
A review of the latest experimental results on charm particle
lifetimes is presented. The most significant update is that the $D_s^+$
lifetime is conclusively larger than the $D^0$ lifetime and signifies
that W-exchange/W-annihilation contributions are large.
Using new high statistics data on $D^+\rightarrow K^+\pi^+\pi^-$
together with the $D_s^+$ lifetime and some assumptions, one can 
phenomenologically extract the strength of the W-exchange contribution
in $D^0$ decays and of W-annihilation in $D_s^+$ decays. These are
larger than or at the limit of theoretical expectations using QCD-based
operator production expansion techniques.
}

\begin{document}

\section{Introduction}

\subsection{Motivation}

The study of charm particle lifetimes is broadly
motivated by two main goals. The first is to enable the conversion of relative
branching fractions to partial decay rates and the second is to
learn more about the strong interaction. 

Experimental data on charm decays
are normally obtained by measuring decay fractions, {\em e.g.}
$\Gamma(D^0\rightarrow K^-\pi^+)/\Gamma_{tot}(D^0)$, 
whereas theory calculates the partial decay rate,
$\Gamma(D^0\rightarrow K^-\pi^+)$. The lifetime of the
particle, $\tau = \hbar/\Gamma_{tot}(D^0)$, is needed in order
to convert the experimentally measured decay fractions into decay
rates. Not only does this allow tests of theoretical predictions
but it also enables the extraction of Standard Model parameters if
the theoretical calculations are reliable, {\em e.g.} a comparison
of $D$ semileptonic decay rates may allow a direct extraction of 
$\vert V_{cs}\vert$ and $\vert V_{cd}\vert$ allowing a test of the
unitarity of the CKM matrix.

The second motivation for the study of lifetimes is that they are
interesting in their own right. They allow us to learn more about
the ``Theoretically-Challenged'' part of the Standard Model,
{\em i.e.} non-perturbative QCD. This is one of the few areas of the 
Standard Model where experimental data and theoretical ideas closely
interact and is thus intellectually interesting. For example, even
though we have some models,
we have little idea about exactly {\em how} quarks turn into hadrons and
we are still learning about the importance of different contributions
to quark decays.
Calculations using Lattice QCD are only just now
being used to study the {\em dynamics} of decays and
reliable results are still being eagerly awaited \cite{askhq98}.

\subsection{Decay Diagrams}

The lifetime of a particle is given by the following expression:
\begin{equation}
\tau={\hbar\over\Gamma_{SL}+\Gamma_{NL}+\Gamma_{PL}}
\end{equation}
where $\Gamma_{SL}$ is the semileptonic decay rate, ({\em e.g.}
$\Gamma(D^+\rightarrow\ell^+\nu_{\ell}X)$), $\Gamma_{NL}$ is the
non-leptonic or hadronic decay rate, ({\em e.g.}
$\Gamma(D^+\rightarrow{\rm hadrons})$), and $\Gamma_{PL}$ is the
purely leptonic decay rate, ({\em e.g.} 
$\Gamma(D^+\rightarrow\ell^+\nu_{\ell})$).
Compared to the total rate, the purely leptonic decay rate is normally 
very small due to helicity suppression.\footnote{The
$D$ meson has spin $0$ so that in the decay,
the resulting lepton
(anti-lepton) and anti-neutrino (neutrino) must {\em both} be either
left-handed or {\em both} right-handed in order to conserve angular
momentum. However the $V-A$ nature of the
weak interaction requires left-handed particles and right-handed
anti-particles\cite{aitchison89}.}
In addition
current data for $D$ meson decays indicate that the semileptonic rates
for $D^+$ and $D^0$ are equal to within at least about 10\%\ if not 
better.\footnote{The semileptonic decay rate is given by the ratio of the
semileptonic branching ratio to the lifetime. Using the world average
values for these compiled by the Particle Data Group \cite{pdg98},
$\Gamma_{SL}(D^+)=(1.071\pm 0.119)\times 10^{-13}$~GeV and
$\Gamma_{SL}(D^0)=(1.067\pm 0.041)\times 10^{-13}$~GeV.}
This means that the large difference between the observed $D^+$ and $D^0$
lifetimes ($\tau(D^+)/\tau(D^0)=2.55\pm 0.04$) is due to a large
difference in the hadronic decay rates for the $D^+$ and the $D^0$.
Thus in contrast to the spectator model \cite{spect75}
which only has the free charm quark decay diagram and predicts
equal $D^+$ and $D^0$ lifetimes, we need to take into account spectator
quark effects. This entails taking into account other decay diagrams
like those in figure \ref{fig1}\ and any interferences between them.

\FIGURE{\epsfig{file=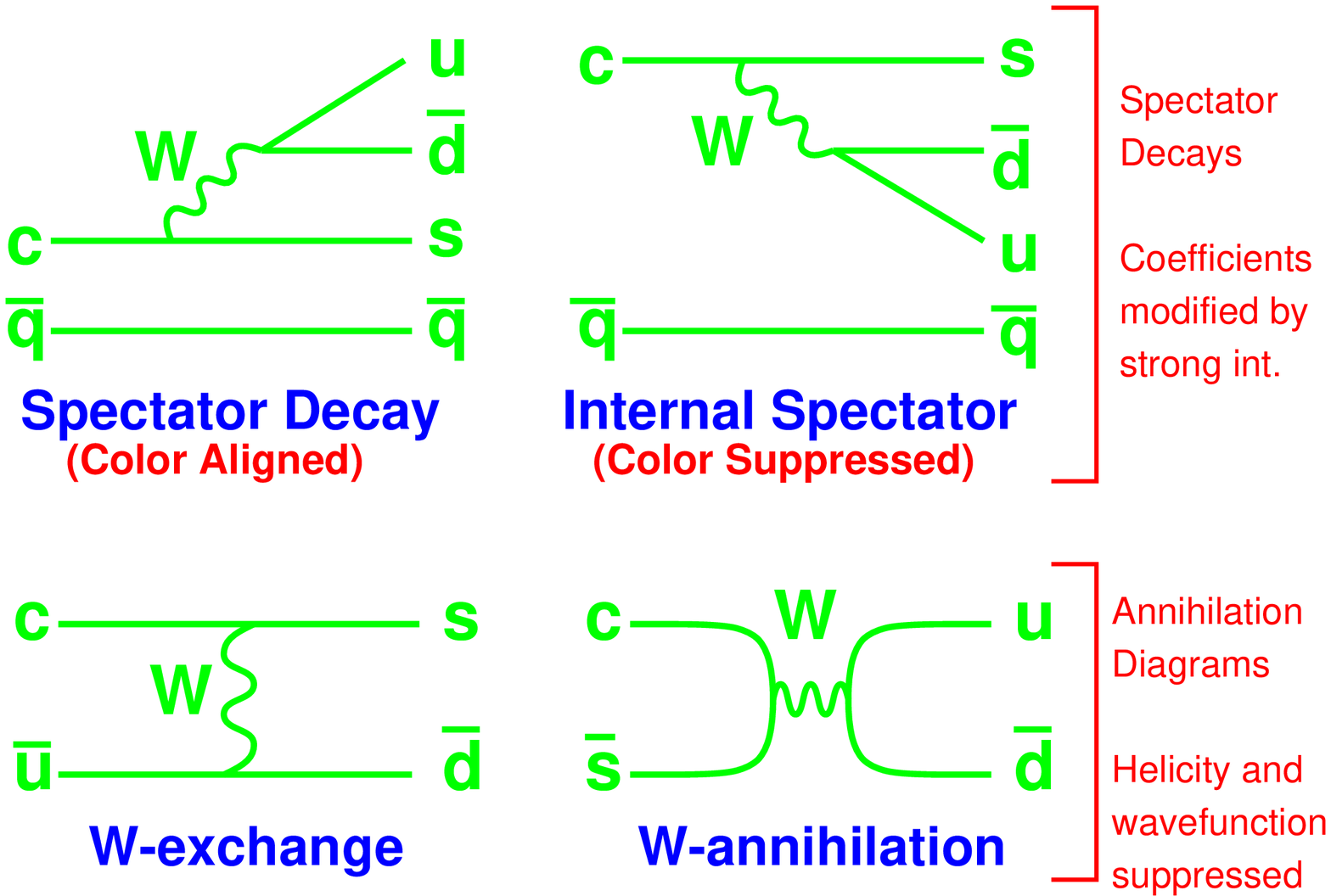,width=7.0cm}
\caption{Hadronic decay diagrams for charm meson decays.}
\label{fig1}}

The conventional wisdom used to explain the smaller hadronic width of
the $D^+$ relative to the $D^0$ is that in the $D^+$ Cabibbo-allowed
decays ($c\bar{d}\rightarrow s(u\bar{d})\bar{d}$), there exist identical
quarks in the final state unlike for $D^0$, so there are additional
(destructive) interference contributions for the $D^+$. Or, we can talk
about a model where one views
the interference as that occurring 
between the external spectator and internal spectator
decay diagrams of figure \ref{fig1} which can lead to the same exclusive
final state. Ignoring the more complicated soft gluonic exchanges,
it is relatively easy in this model 
to roughly show that the additional interference for
inclusive hadronic decays for $D^+$ is destructive and can lead to a lifetime
ratio of $\tau(D^+)/\tau(D^0)\sim 2.0$. However it is difficult to determine
exactly how large a ratio of $\tau(D^+)/\tau(D^0)$ interference
effects can accommodate and therefore how large is the additional contribution
of Cabibbo-allowed W-exchange decays needed for the $D^0$. 
One has to take care in calculating the size of the Pauli interference since
naive calculations can produce too large a value resulting in a negative
total decay rate for the $D^+$ \cite{hycheng99b}.
Cabibbo-allowed W-exchange
decay is expected to contribute to lowering the $D^0$ lifetime but this
contribution is wavefunction and helicity suppressed 
($\sim \vert f_d\vert^2m_s^2/m_c^4\sim 10^{-3}$ without gluon
exchange) and is difficult to calculate reliably.

Clearly a better understanding of charm inclusive
decays is necessary. Experimental data on lifetimes from all the
charm particles will allow us to learn more about how they
decay and in turn use the data to extract Standard Model parameters
like quark masses and the CKM matrix elements $\vert V_{cs}\vert$
and $\vert V_{cd}\vert$.

\subsection{Theoretical Overview}

A systematic approach now exists for the treatment of
inclusive decays that is based on QCD and consists of
an operator product expansion in the Heavy Quark Mass \cite{bdecays}.
In this approach the decay rate is given by:
\begin{equation}
\Gamma_{H_Q}={G_F^2m_Q^5\over 192\pi^3}\Sigma f_i\vert V_{Qq_i}\vert^2
\left[A_1+{A_2\over\Delta^2}+{A_3\over\Delta^3}+...
\right]
\end{equation}
where the expansion parameter $\Delta$ is often taken as the heavy quark
mass and $f_i$ is a phase space factor. 
$A_1=1$ gives the spectator model term and the $A_2$ term produces
differences between the baryon and meson lifetimes. The $A_3$ term includes
the non-spectator W-annihilation and Pauli interference effects. 
For meson decays, parts of
these terms can be related to certain observables whereas 
for baryons one relies solely on particular
quark models or QCD sum rules to determine the parameters fully.
The importance of higher order terms is not really known though some
studies have pointed to possibly large higher order 
contributions \cite{hycheng99b,blokshifman93}.

A theoretical review is outside the scope of this article and the reader is
referred to other reviews \cite{bdecays}.

\section{Review of Experimental Results}

There have been new measurements of charm lifetimes
since the 1998 review performed by the PDG \cite{pdg98}. Some are
results published in journals while others were presented at
conferences this year. Table \ref{tabexpt}\ shows the experiments
that have shown new lifetime measurements.

\TABULAR{ll}{
\hline
{\bf Experiment}& {\bf Beam Type} \\
\hline
E791 & 500 GeV $\pi^-$\\
CLEO & $e^+e^-$ collider at $\Upsilon(4S)$ \\
FOCUS& 190 GeV photon \\
SELEX& 600 GeV $\Sigma^-$ and $\pi^-$\\
\hline
}{Charm experiments\label{tabexpt}}
\subsection{Experimental Method}

\TABULAR{lcccc}{
\hline
{\bf Experiment}& {\boldmath $\tau(D^+)$} {\bf fs} &
{\boldmath $\tau(D^0)$} {\bf fs} &
{\boldmath $\tau(D_s^+)$} {\bf fs} &
{\boldmath $\tau(\Lambda_c^+)$} {\bf fs} \\
\hline
PDG98&     $1057\pm 15$ & $415 \pm 4$      & $467 \pm 17$ & $206\pm 12$\\
E791$^a$ & $1065\pm 48$ & $413 \pm 3\pm 4$ & $518 \pm 14\pm 7$  & \\
CLEO & $1033.6\pm 22.1^{+9.9}_{-12.7}$ & 
                      $408.5 \pm 4.1^{+3.5}_{-3.4}$ &
                      $486.3 \pm 15.0^{+4.9}_{-5.1}$  & \\
FOCUS$^b$ & &
             & $506 \pm 8$ & $204.5\pm 3.4$ \\
SELEX$^b$ & & & & $177\pm 10$ \\
\hline \hline
World Average& $1052\pm 12$ & $412.8\pm 2.7$ & $499.9\pm 6.1$ & 
        $201.9\pm 3.1$\\
\hline
\multicolumn{5}{l}{\small $a$ $\tau(D^+)$ using only the $\phi\pi^+$ mode, no
systematic uncertainty quoted} \\
\multicolumn{5}{l}{\small $b$ Preliminary result with no 
systematic uncertainty quoted} \\
}{Summary of new charm lifetime measurements split by experiment\label{tab1}}

Unlike the lifetime measurements for the $b$ particles, the methods used
for measurements of the charm particle lifetimes are more straight forward.
Firstly, the number of reconstructed charm decays are large enough that
only exclusive decays are used -- inclusive methods are not needed. This 
means that the charm particle momentum is fully measured.

For the fixed target experiments, the resolutions of the production and
decay vertices are about 10~$\mu$m in each of the two directions
transverse to the beam direction and about 400--600$\mu$m along the
beam direction. The resolution varies with the multiplicity of
charged tracks in the vertices as well as on the momenta of the charged tracks. 
Since the boost is typically large 
($\langle\beta\gamma\rangle\sim 40$--$100$) 
the full 3-dimensional decay length ($\ell$) is
used to measure the proper time for the decay,
$t=\ell/\gamma\beta c = (\ell/c)\times (m_D/p_D)$, where $p_D$ and $m_D$ 
are the
momentum and rest mass of the charm particle respectively. The typical
proper time resolution is about 40--60~fs for E791 and FOCUS and
is smaller, $\sim$20~fs, for SELEX due to their much larger average
$D$ momentum. To eliminate background, charm candidates are selected that
have a large separation between the production and decay vertices,
typically by
many $\sigma_{\ell}$, {\em i.e.}
$\ell>N\sigma_{\ell}$. This selection drastically reduces
the acceptance of candidates with short lifetimes and the acceptance as
a function of proper time is rapidly varying at short proper times.
In order to reduce the systematic uncertainty that would be associated with
having to know this acceptance function accurately, one uses the reduced
proper time, $t^{\prime}=t-(N\sigma_{\ell}/c)\times(m_D/p_D)$. The acceptance
as a function of $t^{\prime}$ is quite flat and therefore only small
acceptance corrections are necessary. 
The effect of using the reduced
proper time is to start the clock at a different point for each charm
candidate event, determined by $\sigma_{\ell}$. One assumes, and can
check that there is
no drastic bias in $\sigma_{\ell}$ that could affect the 
$t^{\prime}$ distribution from following a pure exponential decay.
Any bias would have to be correctly simulated in the Monte Carlo.

Even with the relatively small boost ($\langle\beta\gamma\rangle\sim 1.7$) for
charm mesons produced in a
$e^+e^-$ collider running at the $\Upsilon(4S)$, data from CLEO-II.5 can
be used to measure lifetimes. This is possible due to a newly installed
silicon vertex detector, which enabled CLEO to obtain a resolution on the
decay vertex of 80--100~$\mu$m in the D flight direction in the $xy$ plane.
This corresponds to relatively poor proper time resolutions of about
140--200~fs, but is however sufficient to competitively 
measure the lifetimes of the
charm mesons as these are longer lived than the charm baryons.
Due to the detector and magnetic field arrangement of CLEO, the decay
length and momentum of the charm meson is measured in the $xy$ plane,
(which is transverse to the beam direction). The inherently smaller
backgrounds in $e^+e^-$ collisions allow selection of charm signals
without any vertex detachment selection criteria. This means that the
the absolute proper time $t=(\ell^{xy}/c)\times(m_D/p_D^{xy})$ can be
used, thus eliminating one contribution to the acceptance uncertainty.
However, the relatively large proper time resolution requires good
knowledge of this resolution including non-Gaussian tails which could
bias the fitted lifetime.
Although the new silicon tracker in CLEO-II.5 has 
enabled them to measure lifetimes to a
precision rivaling the fixed target-dominated world averages, the
next generation fixed target experiment FOCUS will be overwhelming
with a huge sample of fully reconstructed charm decays.

The lifetimes are usually extracted using a maximum likelihood fit.
Either a binned (proper time) likelihood or an unbinned 
(candidate-by-candidate)
likelihood is used. For the binned likelihood, events are taken from
the mass peak region with events from mass sidebands giving an estimate of
the background lifetime distribution. For the unbinned
likelihood, the mass as well as the proper time for each charm candidate
is used where candidates from a wide mass region are selected.
As well as fitting for the lifetime, the fraction of background is also
usually varied in the fits. The details of each fit are different for
each lifetime measurement.

\subsection{Measurements of Charm Lifetimes}

The world average lifetimes for the weakly decaying charm particles are
dominated by measurements from Fermilab E687 published in 1993-1995.
These are beginning to be superseded by updates this year to 
the $D$ meson lifetimes as well as to the $\Lambda_c^+$ lifetime.

The CLEO collaboration has published their measurements for the lifetimes of
the $D^+$, $D^0$ and $D_s^+$ \cite{cleoltimes}. The modes
used were $D^0\rightarrow K^-\pi^+$, $K^-\pi^+\pi^0$, $K^-\pi^+\pi^-\pi^+$,
$D^+\rightarrow K^-\pi^+\pi^+$, and $D_s^+\rightarrow\phi\pi^+$ with
$\phi\rightarrow K^+K^-$. Besides the usual vertexing requirements,
to additionally suppress backgrounds they required that the $D^0$ and $D^+$
come from $D^{\ast +}$ decays to $D^0\pi^+$ and $D^+\pi^0$ respectively.
The momentum of the $\pi^0$ in the decay $D^0\rightarrow K^-\pi^+\pi^0$
is required to be $>$ 100~MeV/c and the $D^{\ast +}$ and $D_s^+$ mesons
are required to have momenta larger than 2.5~GeV/c. A seven parameter fit is
used to extract the lifetime for each mode before any averaging
is done. Three proper
time resolutions are used in the fit, two of them to model underestimates of
the 
mismeasurement errors. Two backgrounds are fitted, one with zero lifetime
and another component with a finite lifetime. An unbinned likelihood is
used but with the probability associated with the candidate mass 
determined in a separate (mass) fit. The CLEO measurements are shown
in table \ref{tab1}, and the figures are available in their
publication \cite{cleoltimes}.

E791 is a hadroproduction experiment that took data in 1990--1991 at
Fermilab and new measurements using these data have recently been
published for the
lifetimes of the $D_s^+$ \cite{e791ltimes1}\  and the $D^0$
\cite{e791ltimes2}. Figures of the signals and lifetime
fits are available in these
publications.
For the $D_s^+\rightarrow\phi\pi^+$ 
measurement, due to the requirement of a resonance $\phi\rightarrow K^+K^-$,
only a loose \v Cerenkov
particle ID requirement is made on the kaon with the same sign as the
pion. However any possible background from $D^+\rightarrow K^-\pi^+\pi^+$
where one of the pions is misidentified as a kaon is eliminated by
removing candidates that have a $K^-\pi^+\pi^+$ mass within $\pm$30 MeV/c$^2$
of the $D^+$ mass. This selection requires that the  background mass
distribution be modeled with a piecewise linear function with a discontinuity
fixed at 1.95 GeV/c$^2$. An unbinned likelihood fit is performed over the
whole mass range that extracts the $D^+$ lifetime as well as the $D_s^+$
lifetime for this mode. In order to reduce any uncertainty in the
acceptance, the acceptance is not
obtained using only
a Monte Carlo simulation, instead
$D^+\rightarrow K^-\pi^+\pi^+$ data are used together with 
the ratio of Monte Carlo
$\phi\pi$ and $K\pi\pi$ acceptances.
The acceptance for $K\pi\pi$ is obtained by dividing the data distribution
by a pure exponential with the world average $D^+$ 
lifetime, $\epsilon_{data}(K\pi\pi)$. The acceptance for $\phi\pi$ is
then given by $\epsilon_{data}(K\pi\pi)\times \left(
\epsilon_{MC}(\phi\pi)/\epsilon_{MC}(K\pi\pi)\right)$ for each
$t^{\prime}$ bin.
The lifetime results are shown in table \ref{tab1}.
Results are also shown in the table for the
$D^0$ lifetime measured using the $K^-\pi^+$ decay mode. This measurement
was performed together with a lifetime measurement in the $K^+K^-$ decay
mode \cite{e791ltimes2}. Here, a different technique was used to extract the
lifetime since the $K^+K^-$ sidebands do not accurately reflect the background
under the $D^0$ mass peak. Events were split into reduced proper time bins
and the number of $D^0$ signal events was found from a mass fit using a
Gaussian with mean and sigma fixed to that obtained in a fit to all events.
A fit to these signal events 
as a function of $t^{\prime}$ using a single exponential 
after particle identification weighting and acceptance corrections
gives the extracted
lifetime.

\FIGURE{\epsfig{file=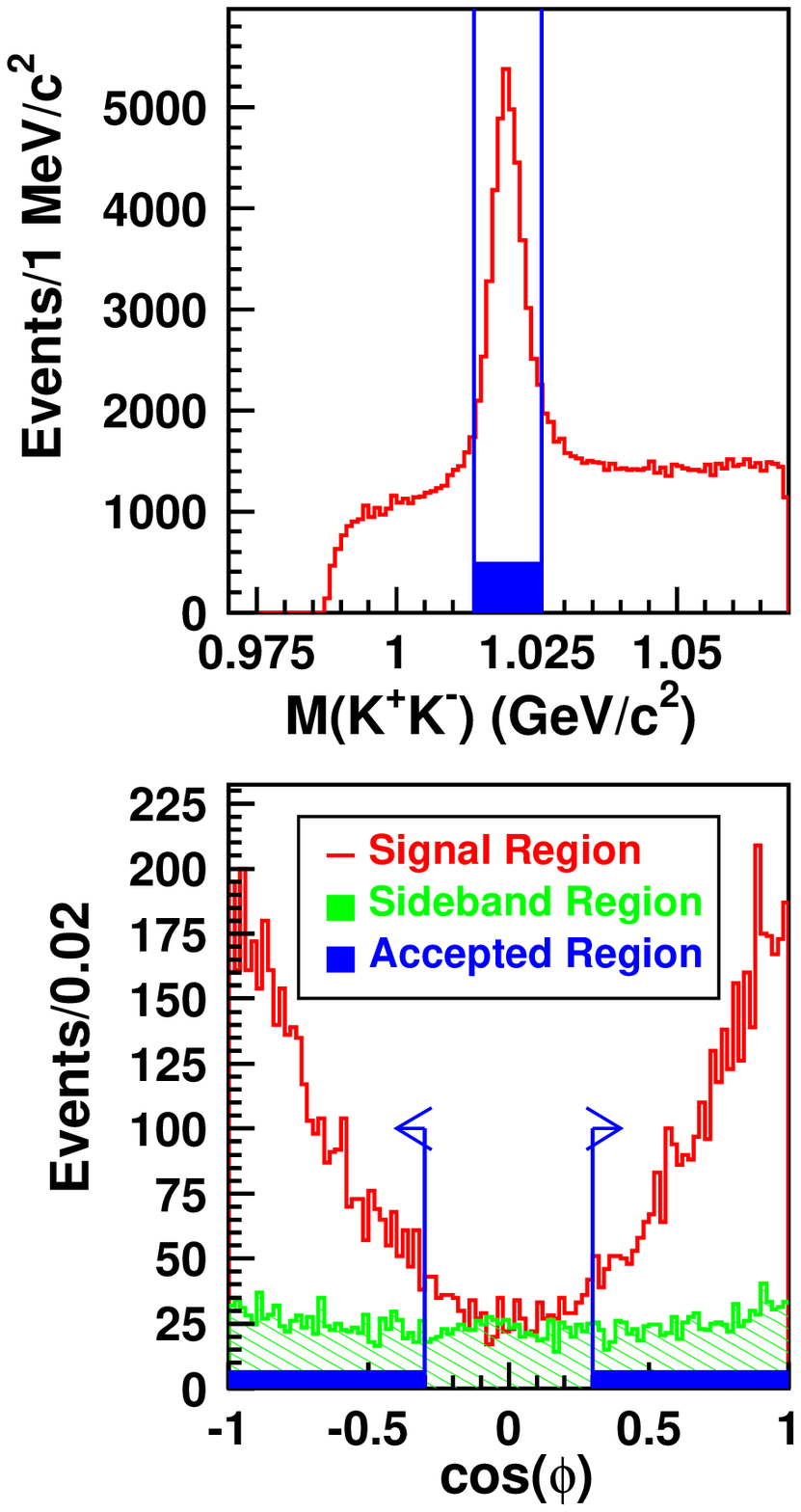,width=3.3cm}
\epsfig{file=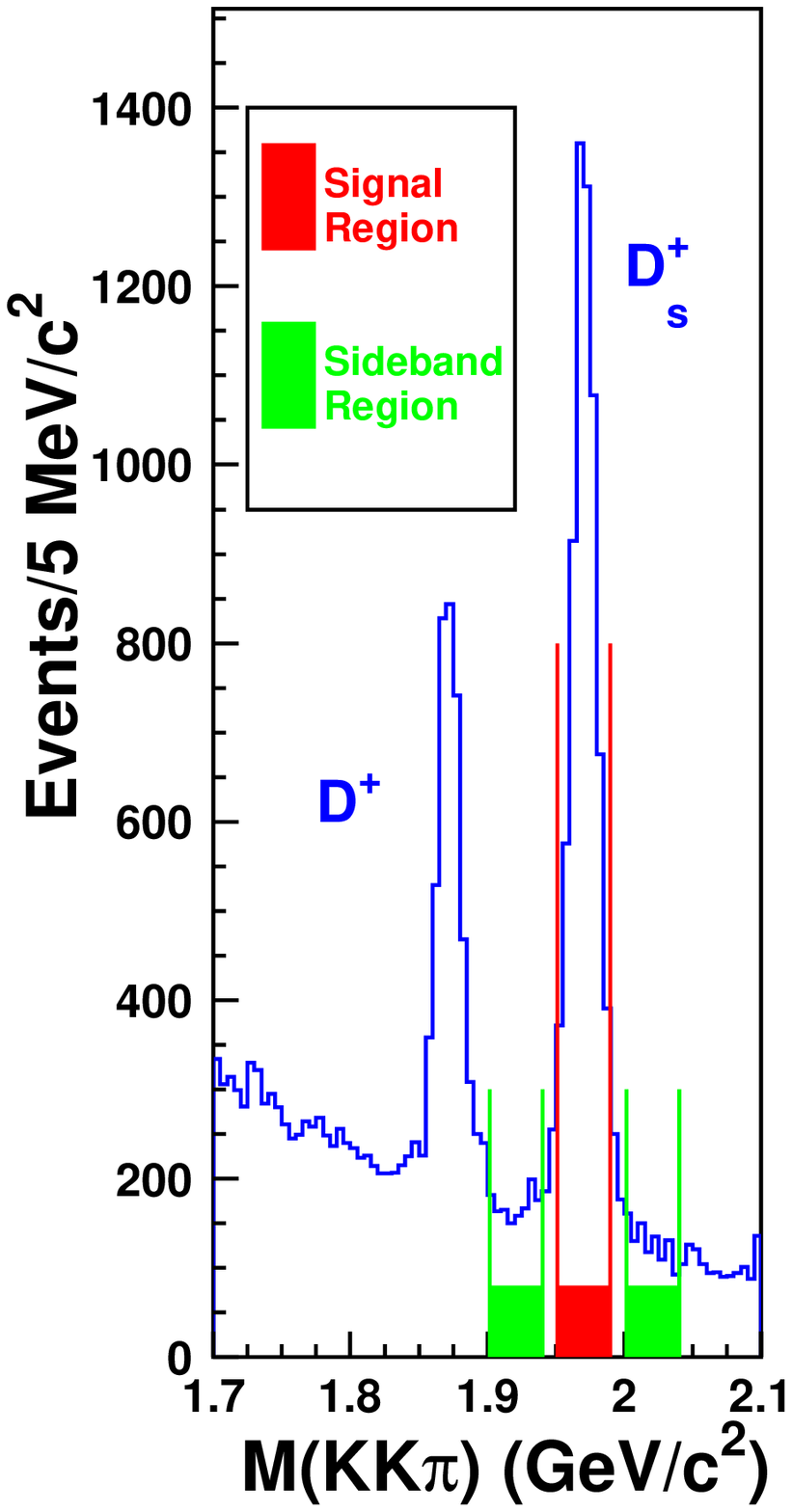,width=3.3cm}
\caption{FOCUS signal for $D_s^+\rightarrow \phi\pi^+$.}
\label{figfocus1}}
\FIGURE{\epsfig{file=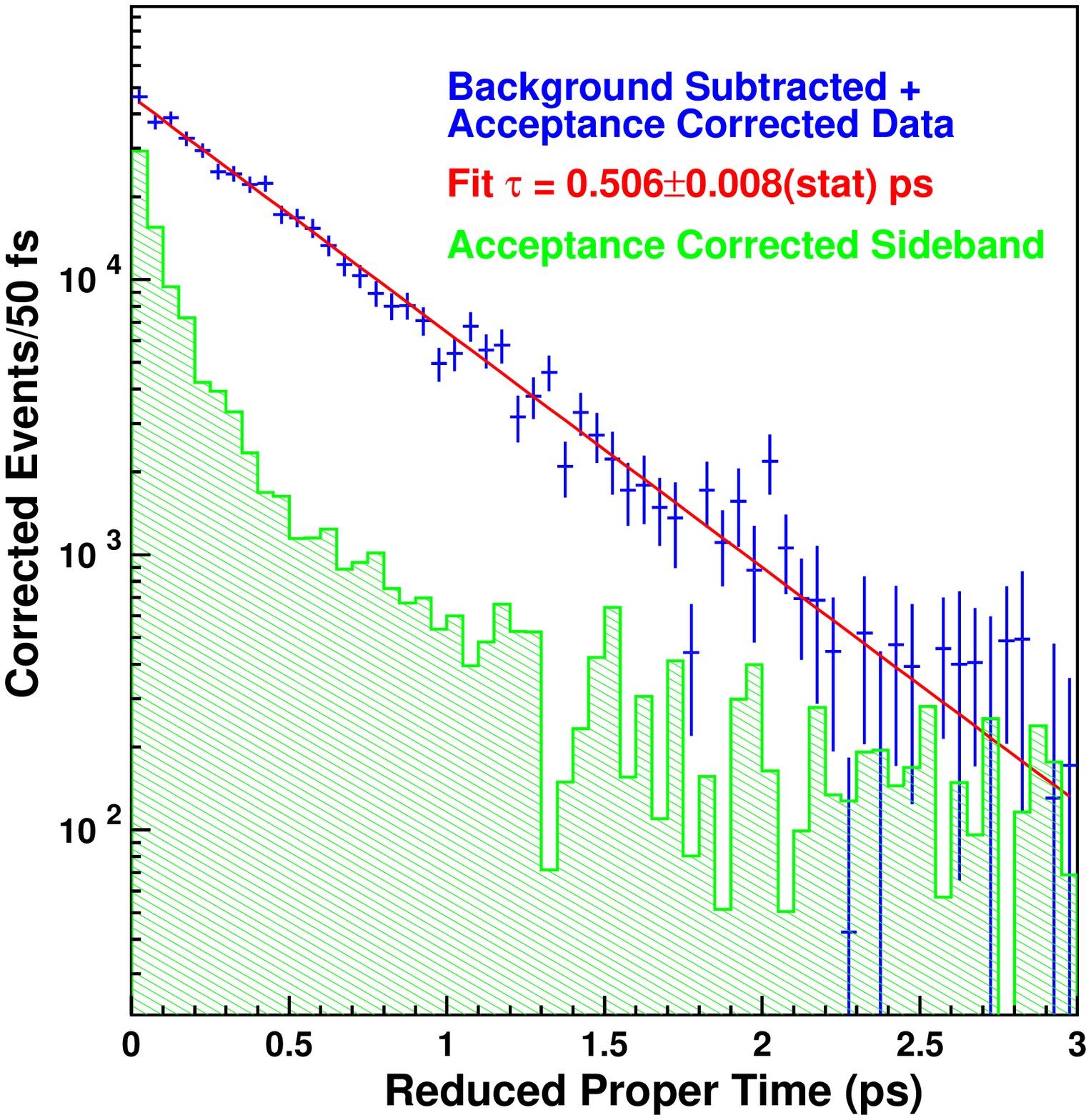,width=7.0cm}
\caption{Preliminary FOCUS lifetime result for $D_s^+\rightarrow \phi\pi^+$.}
\label{figfocus2}}

The Fermilab FOCUS photoproduction experiment took data in 1996--1997 and 
is the follow-on experiment to E687 with
significant improvements to the data quality as well as having collected charm
samples 15--20 times larger than the E687 sample \cite{focus_1}.
A preliminary measurement of the $D_s^+$ lifetime has been made using
50\%\ of the data sample in the decay mode $D_s^+\rightarrow\phi\pi^+$
\cite{focus_2}. The signal and selection regions
are shown in figure \ref{figfocus1}. As well as a cut on the $K^+K^-$
mass to select a $\phi$, a cut is also made on the helicity angle of the
decay. Since the $D_s^+$ and $\pi^+$ each have spin $0$ and the $\phi$
has spin $1$, to conserve angular momentum the $\phi$ and $\pi^+$ must
be in an orbital angular momentum $L=1$ state. Hence the distribution of
the angle between the $\pi^+$ and one of the kaons in the $\phi$
centre-of-mass should vary as $(Y_{L=1}^{m=0})^2\propto cos^2\varphi$
for signal candidates whereas the yield of background candidates are expected
to be independent of $\varphi$. This allows a selection for candidates
with $\vert \varphi\vert>0.3$ to increase signal-to-noise. 
The result of the lifetime fit is shown in
figure \ref{figfocus2}. The preliminary result on the measured lifetime 
using 50\%\ of the FOCUS data is given in table \ref{tab1}.
Also shown in table \ref{tab1}\
is the preliminary measurement of the lifetime of the 
$\Lambda_c^+$ using 80\%\ of the FOCUS data. The $\Lambda_c^+$ is
reconstructed using the $pK^-\pi^+$ decay mode and the signal and
results of the lifetime fit are shown in figure \ref{figfocus3}.
For both
measurements a binned likelihood is used, taking events from the sidebands
as the model for the lifetime distribution for background events
under the charm mass peak. The acceptance is taken from Monte Carlo
simulations. The acceptance correction is small, being larger for
$D_s^+$ than for $\Lambda_c^+$.
 
\FIGURE{\epsfig{file=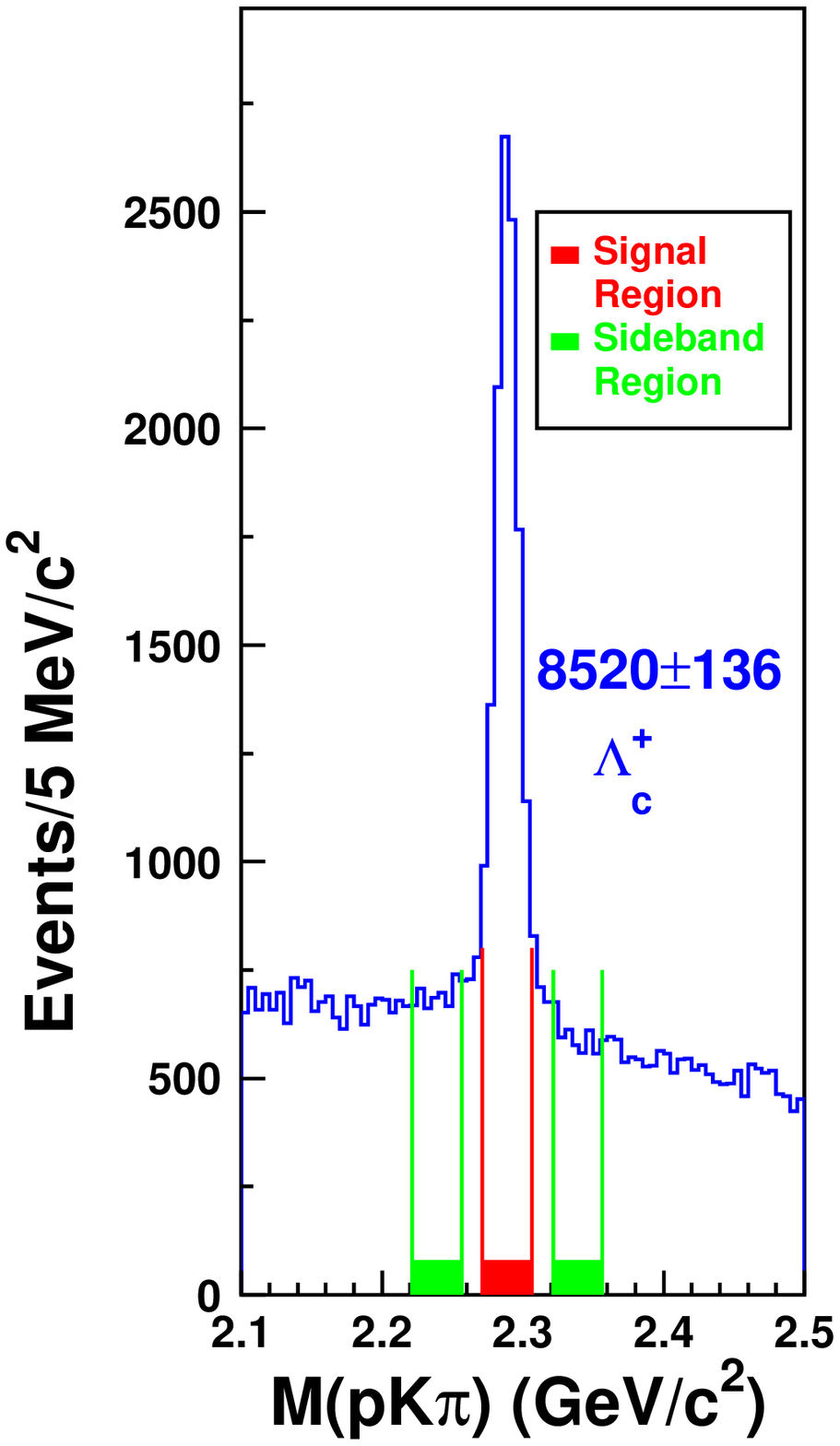,width=2.5cm,height=7.0cm}
\epsfig{file=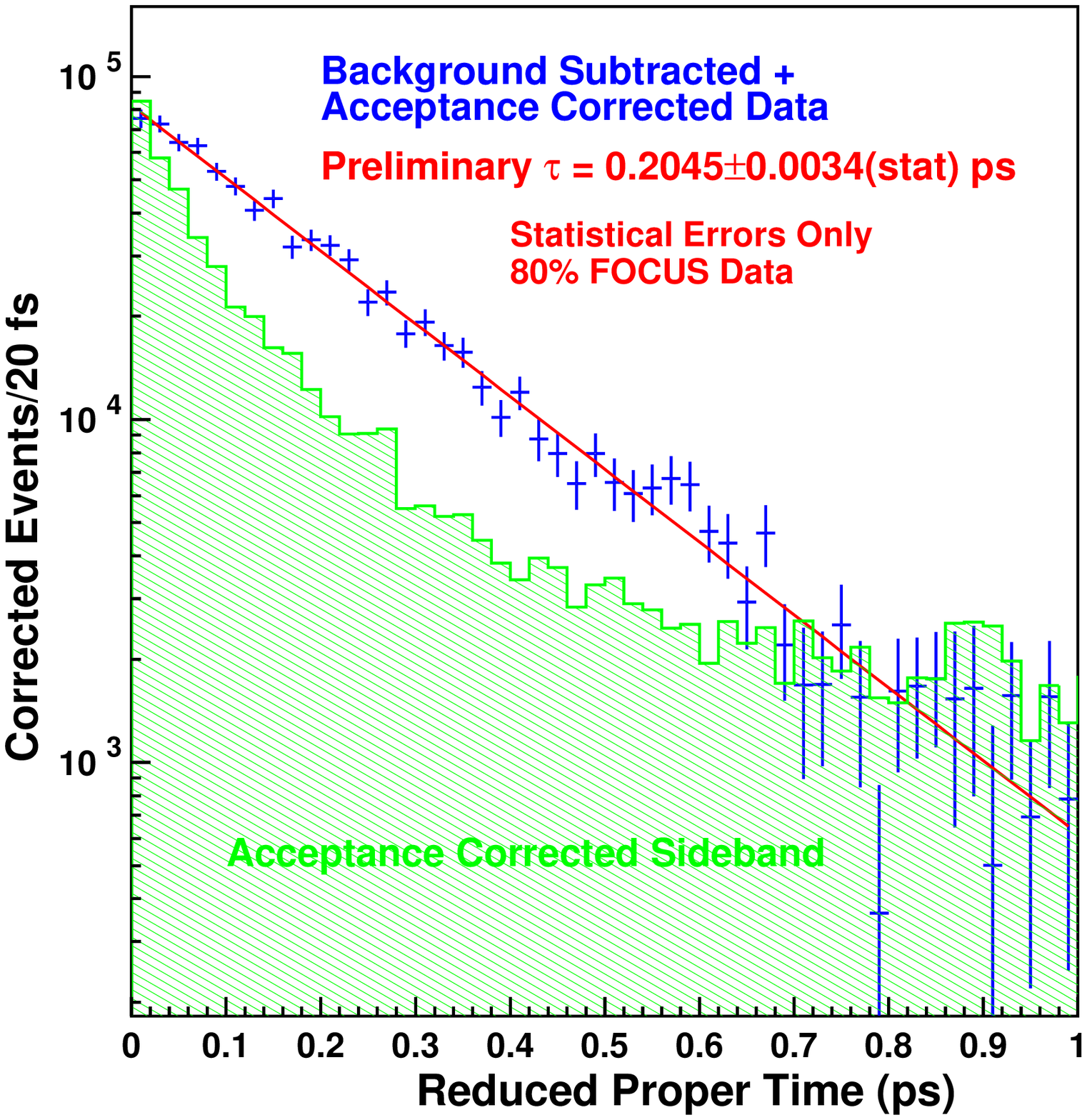,width=4.0cm,height=7.0cm}
\caption{Preliminary signal and lifetime 
results for FOCUS $\Lambda_c^+\rightarrow pK^-\pi^+$.}
\label{figfocus3}}

SELEX is another Fermilab experiment that collected data in 1996-1997.
The data were taken using a 600 GeV $\Sigma^-$ beam and a $\pi^-$ beam.
The experiment was designed for good acceptance in the forward region and
to produce larger fractions of charm-strange baryons. Shown in table
\ref{tab1}\ is a
preliminary measurement of the $\Lambda_c^+$ lifetime using 100\%\ of the
SELEX data in the $\Lambda_c^+\rightarrow pK^-\pi^+$ mode \cite{selexltimes}. 
The acceptance correction was obtained using $D^0$ data and checked
with $K_s^0$ decays which occur near the interaction region. The signal 
and fit are published elsewhere \cite{selexltimes}.

\FIGURE{\epsfig{file=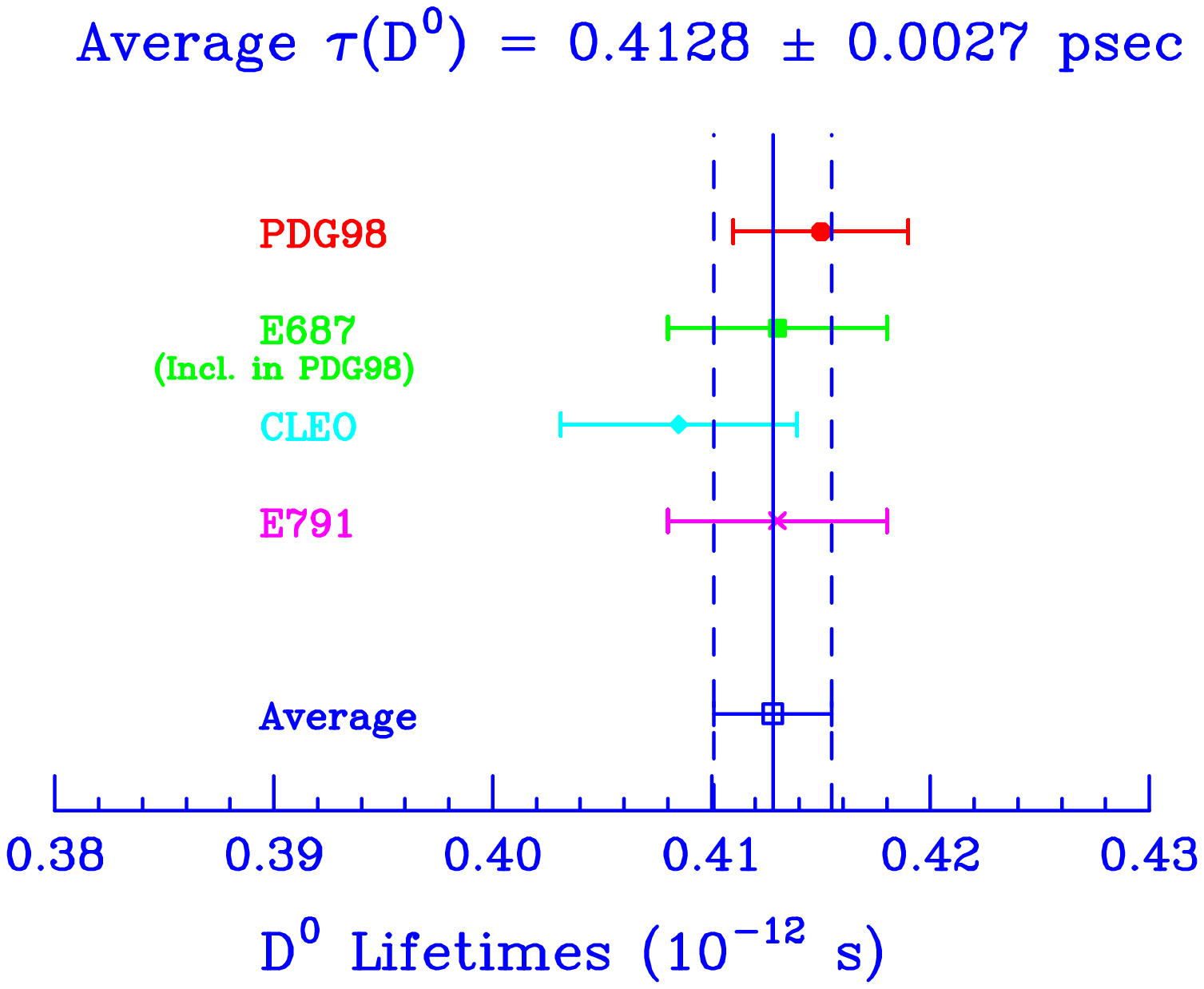,width=5.0cm}
\epsfig{file=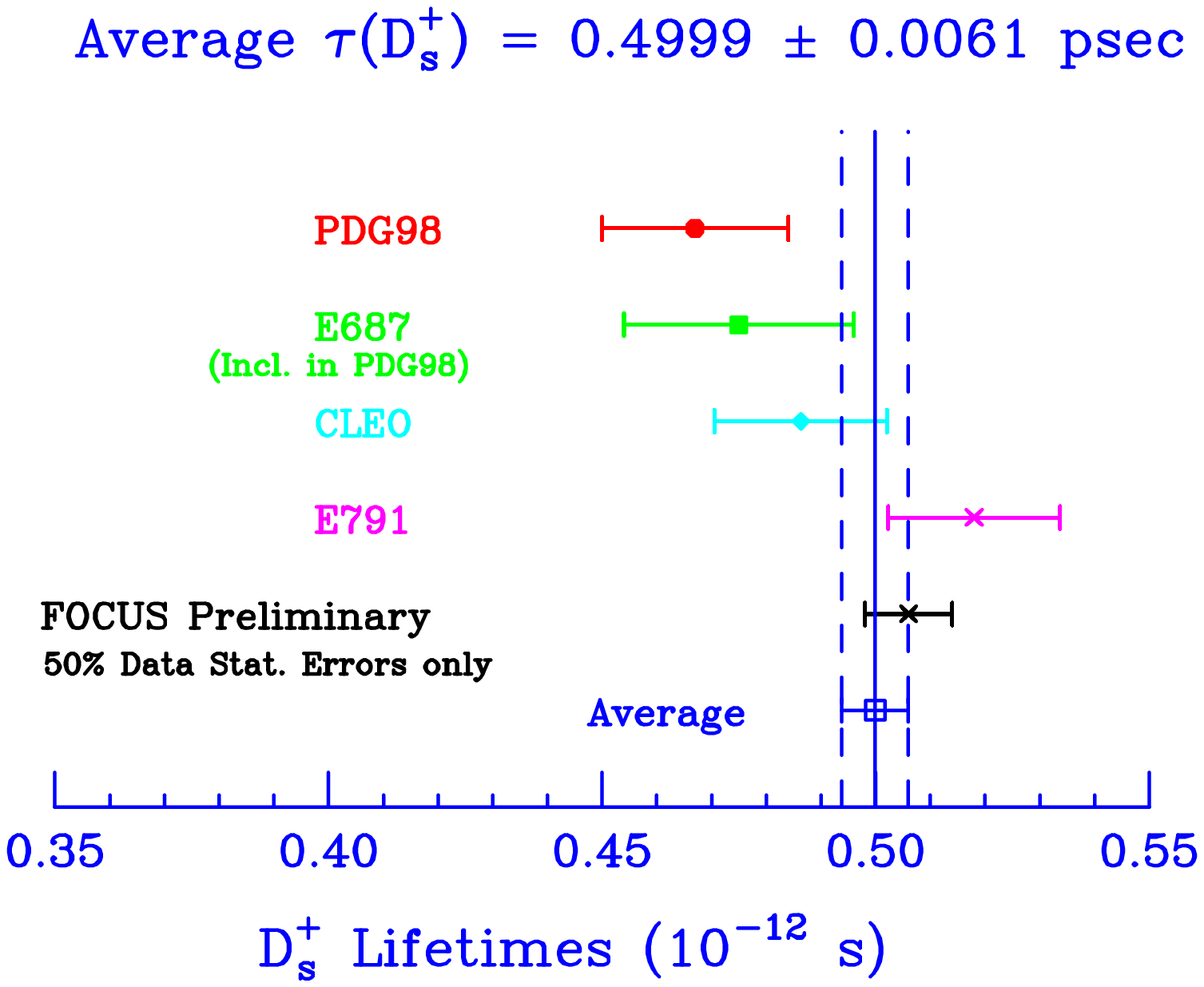,width=5.0cm}
\epsfig{file=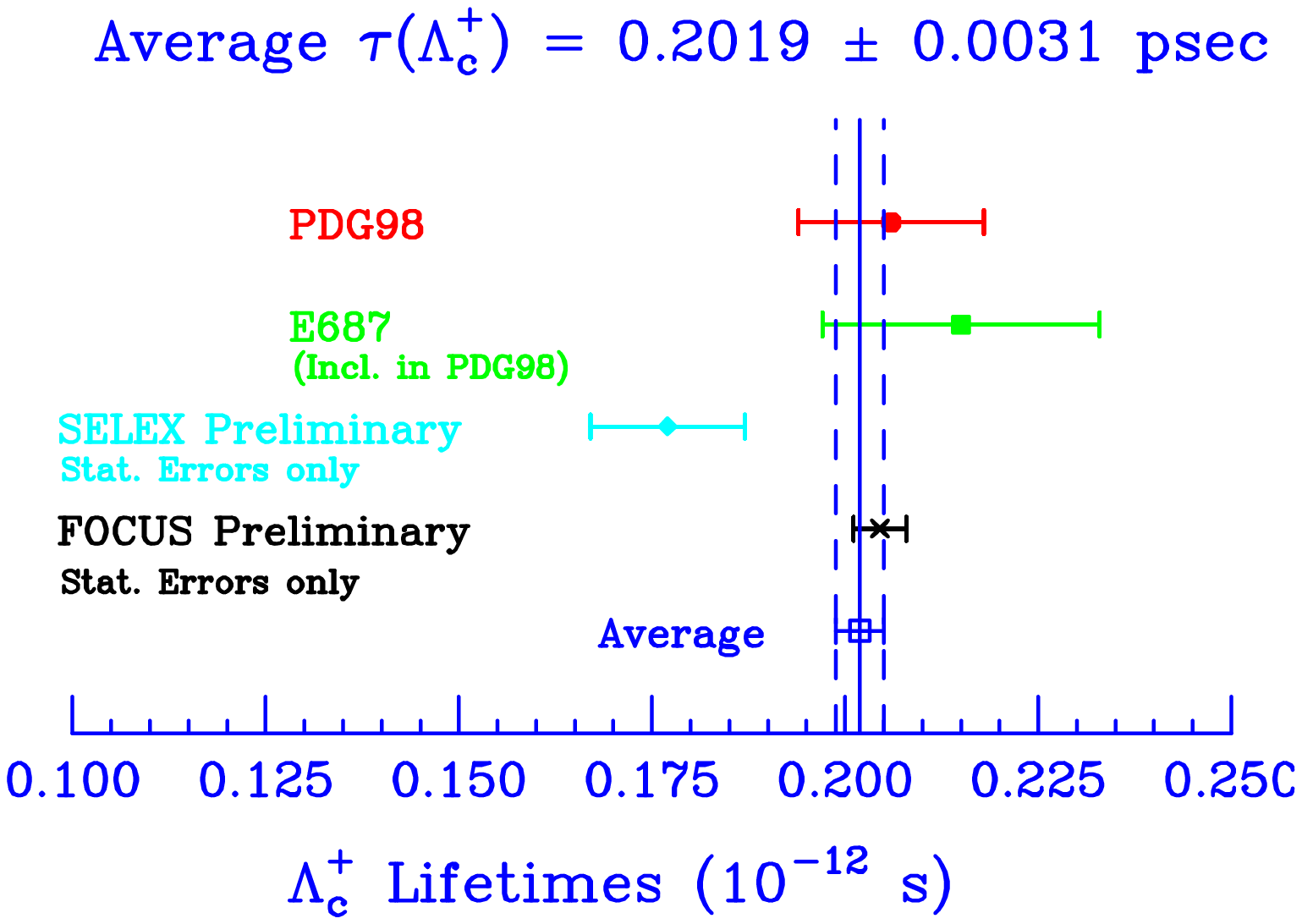,width=5.0cm}
\caption{Summary of new charm lifetime measurements.}
\label{figall}}

The measurements and new world averages are shown in figure \ref{figall}.
The most significant result of these new measurements is that 
$\tau(D_s^+)$
is conclusively larger than $\tau(D^0)$. The world average is now
$\tau(D_s^+)/\tau(D^0)=1.211\pm 0.017$ 
using the FOCUS measurement with
statistical error only, this can be
compared to the earlier PDG98 value of $1.125\pm 0.042$.

\section{Lifetimes and Theory}

\subsection{\boldmath $D^0$ and $D_s^+$ Lifetimes}

The $D_s^+$ lifetime is now conclusively measured to be above the
$D^0$ lifetime, $\tau(D_s^+)/\tau(D^0)=1.211\pm 0.017$. 
Bigi and Uraltsev have used the QCD-based operator product expansion
method
to analyze this lifetime difference and have concluded that 
$\tau(D_s^+)/\tau(D^0)=1.00$--$1.07$ 
is possible without W-annihilation or W-exchange
contributions \cite{bigi_2}.  
The $D_s^+$ lifetime is
reduced by $\sim 3$\%\ due to $D_s^+\rightarrow \ell^+\nu_{\ell}$;
Pauli interference in Cabibbo-suppressed $D_s^+$ decays increase the
$D_s^+$ lifetime by $\sim 4$\%\ ; and $SU(3)_f$ breaking in the
``Fermi motion'' of the $c$ quark is expected to increase the
$D_s^+$ lifetime by $\sim 4$\%, (one can view the quarks in the $D_s^+$
as more confined since $f_{D_s}$ and hence the wavefunction at the original
is larger for $D_s^+$ than for $D^0$.) Any difference in
the measured $D_s^+$ and $D^0$ lifetimes larger than
7\%\ must be attributed to sizable W-annihilation or W-exchange 
(WA/WX) effects.

With their estimation
of the WA/WX contribution, Bigi and Uraltsev conclude that the ratio
$\tau(D_s^+)/\tau(D^0)=1.00$--$1.27$, though $0.8$--$1.27$ is possible
since the sign could change when one allows for interference between the
WA/WX and the spectator contributions \cite{bigi_2}.

In a recent paper Cheng and Yang have also examined the $D_s^+$ and
$D^0$ lifetime difference using the
QCD-based operator product expansion
technique together with the QCD sum rule approach to estimate the hadronic
matrix elements \cite{hycheng99b}. They obtained
$\tau(D_s^+)/\tau(D^0)\approx 1.08\pm 0.04$ including their estimation
of WX/WA contributions to both $D^0$ and $D_s^+$ decays. 
For the size of the WX/WA
they calculate $\Gamma_{WX}(D^0)/\Gamma_{NL}^{Spect}=
0.10\pm 0.06$ and $\Gamma_{WA}(D_s^+)/\Gamma_{NL}^{Spect}=0.04\pm 0.03$,
where $\Gamma_{NL}^{Spect}$ is the spectator decay contribution to the
non-leptonic width.

\subsection{\boldmath Phenomenological Extraction of the 
W-exchange/W-annihilation in 
Inclusive $D^0$ and $D_s^+$ Decays}

With the currently available large charm samples and more precise measurements
of rare branching fractions, one may be able to do more phenomenological
extractions from the data. As an illustration, 
a phenomenological extraction of the strength of the
W-exchange/W-annihilation contribution to
inclusive $D^0$ and $D_s^+$ decays can be done using
some simple assumptions. The
extraction is made possible by a now fairly precise measured value for
\begin{equation}
r_{DCSD}={\Gamma(D^+\rightarrow K^+\pi^-\pi^+)\over
\Gamma(D^+\rightarrow K^-\pi^+\pi^+)}=(6.8\pm 0.9)\times 10^{-3}
\end{equation}
This is an average of the value obtained by the PDG \cite{pdg98}\ and
a preliminary FOCUS measurement of $r_{DCSD}=(6.5\pm 1.1)\times 10^{-3}$
which includes statistical errors only \cite{jew1}.

I make the following assumptions:

\begin{enumerate}
\item
$\Gamma(D^+\rightarrow K^+\pi^-\pi^+)\propto
tan^4\theta_c\cdot\Gamma_{NL}^{Spect}+\Gamma_{WA}^{D^+}$;
\item
$\Gamma(D^+\rightarrow K^-\pi^+\pi^+)\propto
\Gamma_{NL}^{PI}$;
\item
$\Gamma_{WA}^{D^+}<< tan^4\theta_c\cdot\Gamma_{NL}^{Spect}$; and
\item
No interference between the WA/WX contribution and the spectator contribution.
\end{enumerate}
Assumptions 1 and 2 make a possibly dubious relationship between an
exclusive decay rate and a part of the inclusive rate. This could be
approximately accurate if the effects of resonances and final state
interactions in these decays are small enough to allow this assumption.
With assumption 3, one can set $\Gamma_{WA}^{D^+}=0$ in assumption 1.
Finally assumption 4 gives $\Gamma_{tot}(D^0)=\Gamma_{NL}^{Spect}+
\Gamma_{SL}+\Gamma_{WX}$ and $\Gamma_{tot}(D^+)=\Gamma_{PI}+\Gamma_{SL}$.

Using $r_{DCSD}=(6.8\pm 0.9)\times 10^{-3}$ together with
$\tau(D^+)/\tau(D^0)=2.55\pm 0.04$ and
$\Gamma_{SL}/\Gamma_{tot}=0.135\pm 0.006$ obtained from the measured value of
$BR_{SL}(D^0\rightarrow X\ell\nu_{\ell})$, the value for the strength of the
W-exchange contribution can be extracted:
\begin{equation}
{\Gamma_{WX}\over\Gamma_{NL}^{Spect}}=0.29\pm 0.17
\end{equation}
where the error is just from the measured quantities and does not of course
include uncertainties implicit in the assumptions of this model.
The error is dominated by the error in $r_{DCSD}$.

In addition, using $\tau(D_s^+)/\tau(D^0)=1.211\pm 0.017$, 
$\Gamma_{NL}^{Spect}(D_s^+)=\alpha\Gamma_{NL}^{Spect}$ and together with
$\Gamma_{WA}(D_s^+)=\beta\Gamma_{WX}$, the relative strength of the
W-annihilation in $D_s^+$ decays to W-exchange in $D^0$ decays can be
extracted to be $\beta = 0.33$ and thus:
\begin{equation}
{\Gamma_{WA}(D_s^+)\over\Gamma_{NL}^{Spect}}=0.10
\end{equation}
The value of $\alpha$ has been taken to be $1/1.07$ to account for the
differences between the $D_s^+$ and $D^0$ non-spectator decay contributions
mentioned in the previous section.

This illustration only serves to give a somewhat more quantitative 
measure of the unexpectedly large size of the W-exchange/W-annihilation
contributions. The phenomenologically extracted values of these
are 2--3 times larger than those calculated by
Cheng and Yang \cite{hycheng99b}. A more detailed model treatment is limited
by the large uncertainties on some of the measured quantities 
used.\footnote{If one sets
$\Gamma_{WA}^{D^+}=tan^4\theta_c\times\Gamma_{WX}^{D^0}$
we would get a non-sensible result of $\Gamma_{WA}^{D^+}=-\Gamma_{NL}^{Spect}$.
A more reasonable assumption may be to set
$\Gamma_{WA}^{D^+}=tan^4\theta_c\times\Gamma_{WA}^{D_s^+}$. However other
problems arise here too, 
either because the assumptions are too simplistic or the
measured quantities are still not yet measured precisely enough for a
more sophisticated model.}
Note that we expect $\Gamma_{WX}^{D^0}$ to be different from
$\Gamma_{WA}^{D_s^+}$ since the former is colour-suppressed whereas the
latter is colour-allowed, but also since this in itself would predict the
wrong sign for this difference, there must be more complicated processes, for
example in the gluon exchanges in the two cases.

\section{Conclusions}

A number of new charm particle lifetime measurements have been published
or were shown at conferences this year.
The most significant update is that the $D_s^+$ lifetime is now
conclusively measured to be above the $D^0$ lifetime. The ratio
$\tau(D_s^+)/\tau(D^0)=1.191\pm 0.024$ using published measurements.
Using the FOCUS preliminary measurement gives
$\tau(D_s^+)/\tau(D^0)=1.211\pm 0.017$.
This lifetime ratio is now large enough for one to conclude that the
W-exchange contribution in $D^0$ decays is large, estimated to be
about 30\%\ of the
non-leptonic spectator contribution using a simple phenomenological model.
The W-exchange contribution appears to be at the limit of or larger than the
values calculated using the QCD-based
operator production expansion techniques. More precise charm data, for
example in semileptonic decays, is needed to extract the size of the
matrix elements used in these techniques to control the weight of
WA/WX in $D$ decays \cite{hycheng99b}.
Note that this is in contrast to studies
of W-exchange contributions in exclusive $D^0$ decays which is always
complicated by final-state interactions,
e.g. $D^0\rightarrow \phi K_s^0$. If the W-exchange
contribution is as large as the lifetime measurements suggest, then it
must appear somewhere in the exclusive decays. However, conclusive
evidence of W-exchange contributions in exclusive $D^0$ decays is still
missing. Where are they?

We can look forward to more precise charm particle lifetimes from the
Fermilab FOCUS and SELEX experiments, for both charm baryons and mesons.
This should ensure continued theoretical interest in 
the physics of charm lifetimes.

\end{document}